\documentclass[showpacs,showkeys,preprintnumbers,ansmath,amssymb]{revtex4}
\usepackage[dvips]{graphicx}
\newcommand{\be}{\begin{equation}}
\newcommand{\ee}{\end{equation}}
\newcommand{\bea}{\begin{eqnarray}}
\newcommand{\eea}{\end{eqnarray}}
\newcommand{\bm}{\begin{mathletters}}
\newcommand{\eml}{\end{mathletters}}

\begin{document}

\title{Microscopic description of $^{252}$Cf cold fission yields}

\author{M. Mirea$^{1}$, D.S. Delion$^{1,2}$ and A. S\u andulescu$^{2,3}$}
\affiliation{
$^{1}$National Institute of Physics and Nuclear Engineering,\\
407 Atomi\c stilor, Bucharest-M\u agurele, 077125 Romania \\
$^{2}$Academy of Romanian Scientists \\
Splaiul Independen\c tei 54, Bucharest, 050085 Romania \\
$^{3}$Institute for Advanced Studies in Physics,\\
Calea Victoriei 129, Bucharest, Romania}

\begin{abstract}
{We investigate the cold fission of $^{252}$Cf within
the two center shell model to compute the potential energy surface.
The fission yields are estimated by using the semiclassical 
penetration approach.
It turns out that the inner cold valley of the total potential energy
is  strongly connected with Z=50 magic number.
The agreement with experimental values is very much improved only by 
considering mass and charge asymmetry degrees of freedom.
Thus, indeed cold fission of $^{252}$Cf is a Sn-like radioactivity, 
related the other two "magic radioactivities", namely $\alpha$-decay and 
heavy-cluster decay, called also Pb-like radioactivity.
This calculation provides the necessary theoretical confidence to
estimate the penetration cross section in producing superheavy nuclei,
by using the inverse fusion process.}
\end{abstract}

\vskip1cm



\maketitle


\rm

\setcounter{equation}{0}
\renewcommand{\theequation}{\arabic{equation}}

\section{Introduction}
\label{sec:intro}

The synthesis of superheavy elements beyond $Z=104$, suggested by Flerov
\cite{Fle69}, was predicted within the so-called fragmentation theory in
Ref. \cite{San76} by using the cold valleys in the potential energy surface
between different combinations, giving the same compound nucleus.
Soon it was shown in Refs. \cite{Gup77,Gup77a} that the most favorable
combinations with $Z\geq 104$ are connected with the so-called Pb potential
valley, i.e. the same valley of the heavy cluster emission \cite{San80}.

Due to the double magicity of $^{48}$Ca, similar with $^{208}$Pb, in Ref.
\cite{Gup77a} it was proposed $^{48}$Ca as a projectile on various
transuranium targets.
The production of many superheavy elements with $Z\leq 118$
(corresponding to the last stable element Cf) during last three decades
was mainly based on this idea \cite{Oga01,Oga04,Hof00,Hof04}.

The shape of the potential energy versus various degrees of freedom
plays an important role in predicting the fission/fusion probability
to create superheavy nuclei. These processes take place
along the so-called "cold valleys" of these surfaces.
The $\alpha$-particle emission is connected with the "lightest"
cold valley in the fragmentation potential surface.
On the other hand the "heaviest" side of the cold valley
is given by the cold fission process, i.e. the emission of
two fragments with similar masses in their ground states.
Between theses limits there is a broad region of cold  heavy cluster decays.
These decays are strongly connected with the $^{108}$Pb-cold valley
and this is the reason why they are also called "magic-radioactivities".

In order to check the validity of various approaches describing
fission/fusion processes it is very important to reproduce 
experimentally measured yields. In this context we mention that
systematic measurements were performed for $^{252}$Cf \cite{ham1,ham2}.
They correspond to various excitation energies of emitted fragments, 
including cold processes (neutronless), driven along the above 
mentioned cold valleys.
The aim of this work is to describe the existing experimental cold 
fission data within a fully microscopic approach, namely the two-center 
shell model. 

In the past $^{252}$Cf was investigated within the double folding 
potential method \cite{sandu96,sandu98} emphasizing the role of
higher deformations in a correct description of the cold fission 
distribution of fragment yields. Therefore, the deformations introduced in our
model take into account approximately the multipole deformations up to
$\lambda$=4.
In another work, based on a macroscopic model by determining the tip 
distances for the exit point from the barrier for ground state deformed 
fragments, it was predicted \cite{gonnen91} that the major contribution
in the yield distribution corresponds to the light fission fragment 
$A_{2}\approx$100.
  
\section{Theoretical background}
\label{sec:theor}

We extend the analysis performed in Ref. \cite{Del07} within a simple 
model, to a more reliable microscopic approach to estimate the 
fission/fusion barrier, based on a new version of 
the Super Asymmetric Two Center Shell Model \cite{Mir98}.

\begin{figure}
\resizebox{0.6\textwidth}{!}{\includegraphics{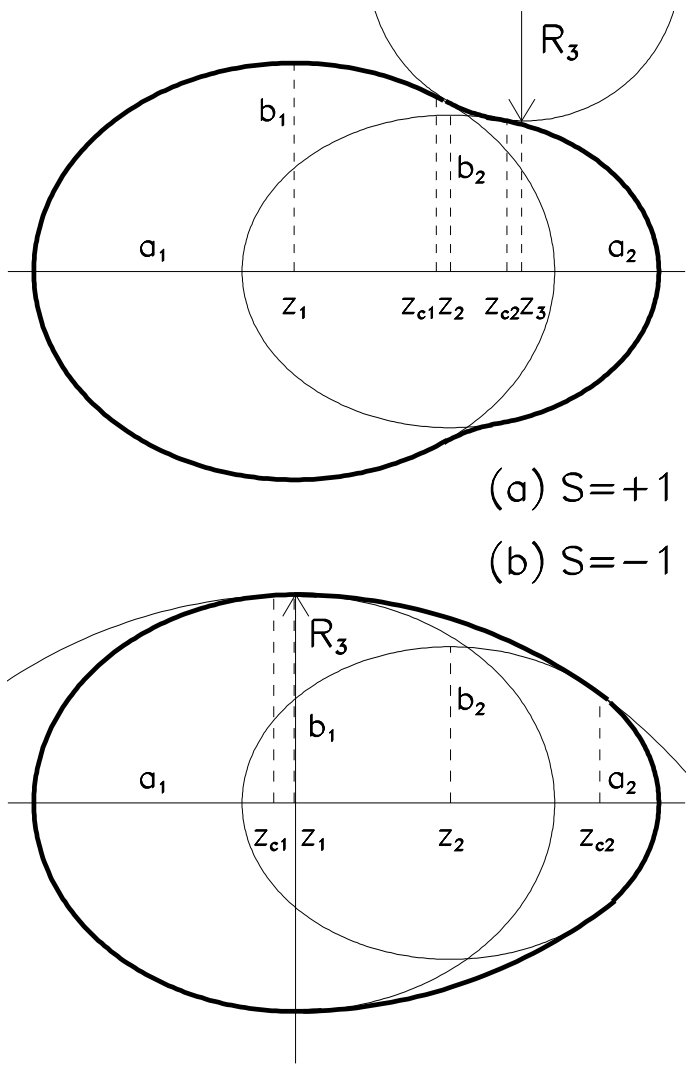}}
\caption{Nuclear shape parametrization.
 }
\label{fig1}
\end{figure}

A nuclear shape parametrization, taking into account the relevant
degrees of freedom encountered in fission, namely the elongation,
the necking, the mass-asymmetry and the deformations of the fragments
is used.
In this context, shell corrections and cranking inertia are realistically
determined.
This model was already used in calculations of some superheavy 
elements \cite{superheavy}. 

\begin{figure}
\resizebox{0.8\textwidth}{!}{\includegraphics{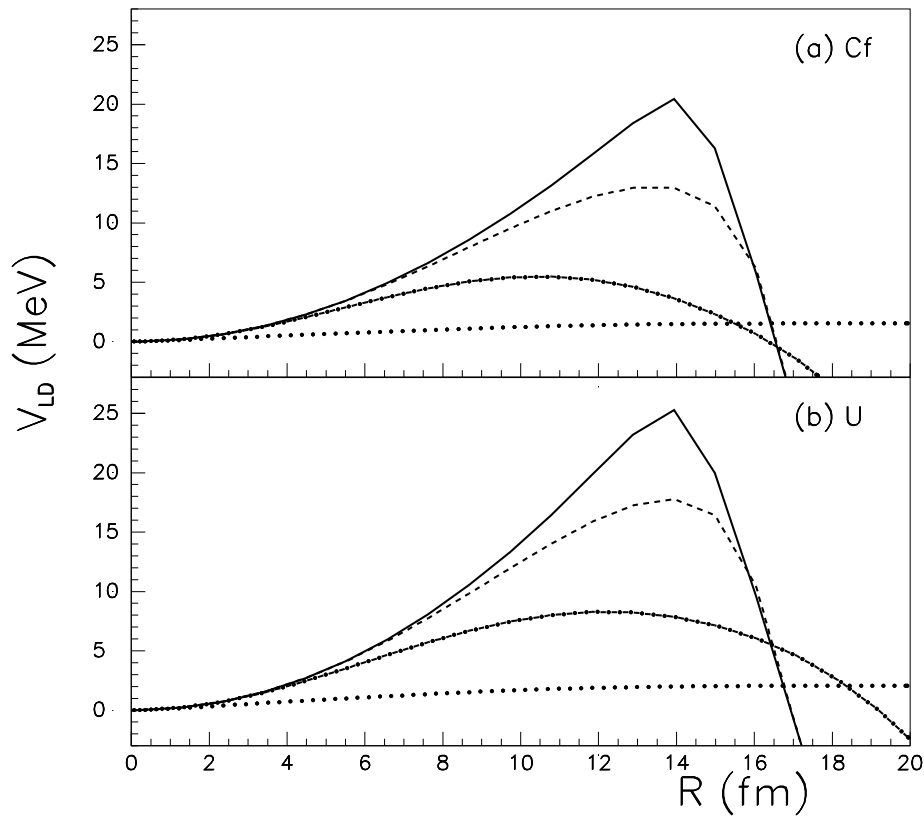}}
\caption{Liquid drop deformation energy as function of the
elongation $R$ for $^{252}$Cf and $^{236}$U as parents. Four values
of the necking parameter are used. $C$=-0.1, 0.1, 0.5 and 100 fm$^{-1}$
corresponding to the dotted, dash-dotted, dashed and full lines, respectively.
In both cases, the light fragment is $^{95}$Rb. 
The parent and the  nascent fragments
are considered spherical. The mass asymmetry is varied linearly from
the initial sphere $\eta$=1 at $R$=0 fm to the final value at the touching configuration
$R=r_{0}(A_1^{1/3}+A_2^{1/3})$ (where $r_0$ is the reduced radius). 
 }
\label{fig2}
\end{figure}

The penetrability, corresponding to some binary partition, defines the
isotopic yield and it is characterized by the difference between the
nuclear plus Coulomb potential and the $Q$-value.
For a given initial nucleus $(Z,A)$ this quantity, called fragmentation 
potential, depends upon the charge, mass numbers of a given fragment (we 
will consider the second one) and the inter-fragment distance.
For a fixed combination $A=A_1+A_2$ the fragmentation potential has a 
minimum at the charge equilibration point $Z_2$, which we will not 
mention in the following, i.e.
\bea
\label{poten}
V(A_2,R)=V_N(A_2,R)+V_C(A_2,R)-Q~,
\eea
where $V_N(A_2,R)$ is the nuclear and $V_C(A_2,R)$ Coulomb 
inter-fragment potential. We also introduced the Q-value in terms
of the difference between binding energies of the parent and
the sum of emitted fragments, i.e.
\bea
\label{Qvalue}
Q=B(Z,A)-B(Z_1,A_1)-B(Z_2,A_2)~.
\eea
For deformed nuclei, due to the fact that the largest emission probability
corresponds to the lowest barrier, the deformation potential decreases 
in the direction of the largest fragment radius.

In most usual treatments of nuclear fission, the whole nuclear
system is characterized by some collective coordinates associated
to some degrees of freedom approximately determining the behavior
of many intrinsic variables. The basic ingredient is the nuclear
shape parametrization.
The generalized coordinates vary in time, leading to the splitting of
the nuclear system.
In this work, the nuclear shape parametrization is obtained by smoothly joining
two spheroids with a third surface given by the rotation of a circle
around the axis of symmetry. This parametrization is characterized
by 5 degrees of freedom, namely the mass asymmetry, the elongation 
$R=z_2-z_1$ given by the distance between the centers of the nascent 
fragments, the  two deformations of the nascent fragments associated to 
the eccentricities $\varepsilon_{i}=\sqrt{1-b_{i}^2/a_{i}^2}$ ($i$=1,2),
and the necking characterized
by the curvature $C=S/R_{3}$ of the median surface and the mass-asymmetry term 
given by the ration of the semi-axis $\eta=a_{1}/a_{2}$. The 
notations used can be identified by inspecting the Fig. \ref{fig1}.
Within this parametrization, swollen shapes in the median surface
can be obtained for a negative curvature ($S$=-1) and necked ones for
a positive curvature ($S$=1). The swollen configurations characterize
the ground states, the shapes during the passage of the first barrier
or the second well, while the necked configurations are associated to
the passage of the second barrier.

In order to estimate the fission yields it is necessary to investigate
the penetration factor. This quantity can be estimated,
as usually, by using the semiclassical integral \cite{Gam28}
\bea
\label{penetr}
P_{A_2}=exp\left\lbrace-2\int_{R_1}^{R_2}
\sqrt{\frac{2\mu(A_2,R)}{\hbar^2}V(A_2,R)} dR\right\rbrace~,
\eea
between internal and external turning points. Two ingredients are
mandatory in order to evaluate the action integral: the 
fragmentation deformation energy $V$ (we will call it simply deformation
energy) and the inertial parameter $\mu$.

In our calculation, the deformation energy of the nuclear system 
is the sum between the liquid drop energy $V_{LD}$ and the shell effects 
$\delta E$, including pairing corrections \cite{nix00}, i.e.
\bea
\label{deform}
V=V_{LD}+\delta E-V_{0}~.
\eea
The energy of the parent nucleus $V_{0}$ is used as a reference
value, so that in the ground state
configuration the deformation energy is zero and asymptotically,
for two separated fragments at infinity, the deformation energy reaches
the minus sum of energies of emitted fragments. Thus, the  above 
Eq. (\ref{deform}) coincides with the definition given by 
Eq. (\ref{poten}).

The macroscopic energy is obtained within the framework of the
Yukawa-plus-exponential model \cite{davies00}
extended for binary systems with different
charge densities \cite{poen1,mm01}. The Strutinsky prescriptions \cite{brac1}
were computed on the basis of a new version of the superasymmetric
two-center shell model. This version solves a Woods-Saxon
potential in terms of the two-center prescriptions as detailed
in Refs. \cite{mirws,mirws2}. The inertial parameter $\mu$ is computed 
in the framework of the cranking model. In some particular evaluations, 
the values of the reduced mass is used.
We considered only cold fission/fusion process. Consequently
the deformations of the initial and final nuclei are given by their ground
state values of Ref. \cite{Mol95}.

\section{Numerical application}
\label{sec:numer}

For comparison with experimental data, the maximal values of the
independent yields for a maximum excitation energy of 7 MeV were
selected from Ref. \cite{ham1}. The selected channels address
binary partitions characterized by the following light fragments: 
$^{95}$Rb, $^{96}$Rb, $^{97}$Sr, $^{98}$Sr, $^{99}$Y, 
$^{100}$Y, $^{101}$Zr, $^{102}$Nb, $^{103}$Zr, 
$^{104}$Nb,  $^{105}$Mo, $^{106}$Nb, $^{107}$Mo, 
$^{108}$Tc, $^{109}$Mo, $^{110}$Tc, $^{111}$Ru, $^{112}$Rh,
$^{113}$Ru, $^{114}$Rh, $^{115}$Rh, $^{116}$Rh, $^{117}$Pd, 
$^{118}$Rh, $^{119}$Pd, $^{120}$Ag, $^{121}$Cd  and $^{122}$Ag. 

\begin{figure}
\resizebox{0.8\textwidth}{!}{\includegraphics{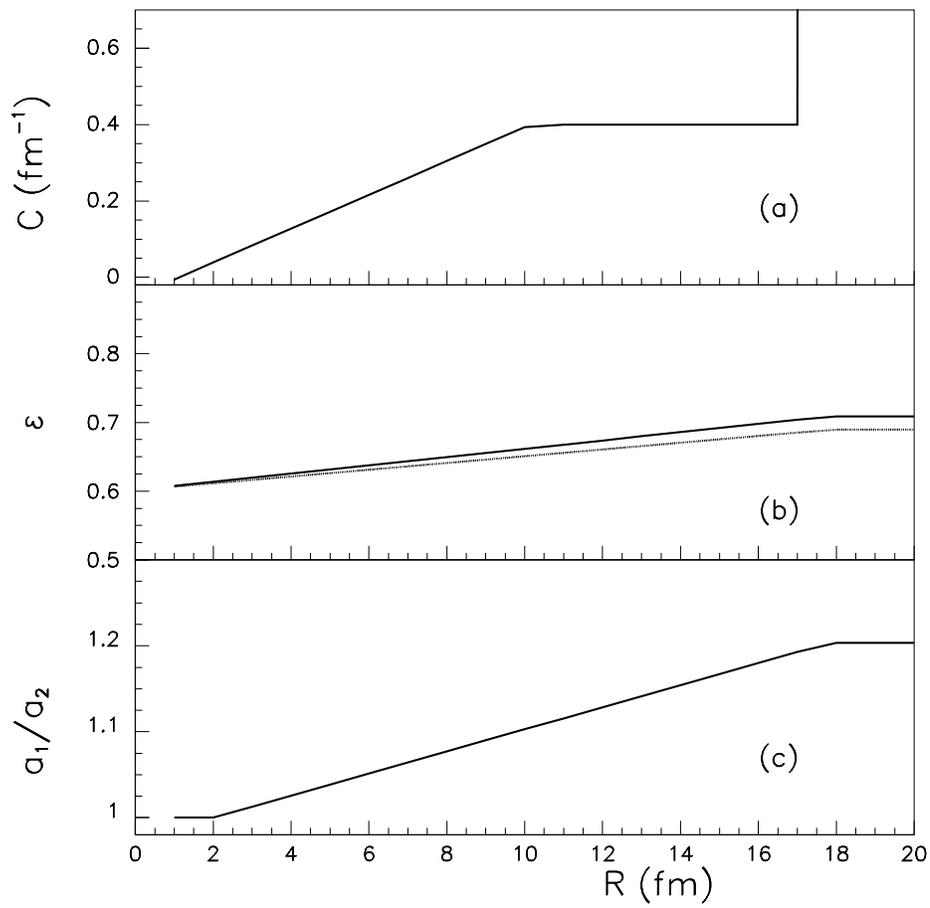}}
\caption{ Variation of the generalized parameters as function of the
elongation. (a)
The necking parameter $C$ varies linearly from the ground state configuration 
up to the region of the first barrier and remains constant up to the
region of the second barrier. In the region of the second
barrier the configurations are necked (that id $C>$0 fm$^-1$), 
and the neck
vanishes at a value close to $R$= 17 fm. The deformations $\epsilon_{1}$
and $\epsilon_{2}$
of the fragments have a linear interpolation between the initial
value of the parent up to the final values of the fragments
at $R$=17 fm. A similar behavior is followed by the 
mass asymmetry parameter $a_{1}/a_{2}$.
 }
\label{fig3}
\end{figure}

\begin{figure}
\resizebox{0.8\textwidth}{!}
{\includegraphics{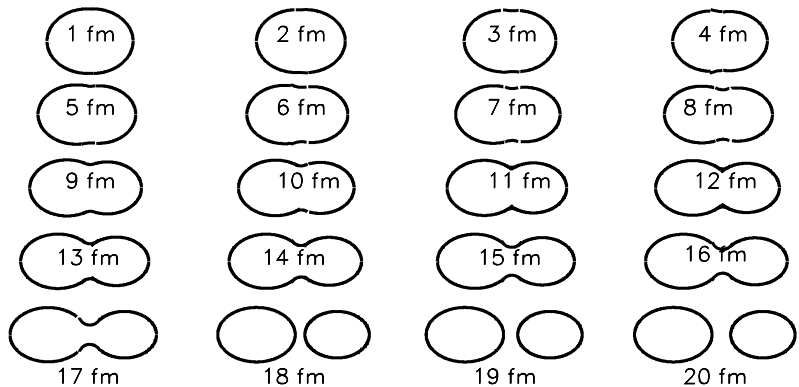}}

\caption{Family of nuclear shapes obtained along the fission path within
the parametrization used in this work. The initial parent nucleus is
$^{252}$Cf and the light fragment is $^{95}$Rb. The values of the
elongation $R$ are marked on the plot. 
 }
\label{fig4}
\end{figure}

First of all, the fission trajectory in our five-dimensional 
configuration space must be supplied, that is a dependence between all 
generalized coordinates. This trajectory starts from the ground-state
of the system and reaches the exit point of the barrier.
The ground-state corresponds to the minimal deformation
energy in the first well.
In order to avoid a complicated determination of the minimal action
trajectory, the deformations of the parent and those of the
two fragments are taken from the literature \cite{Mol95}. A linear variation
from initial values of the nascent fragments eccentricities and of the
mass-asymmetry is postulated starting from the ground state values of 
the parent up to the final ones, characterizing the fragments at the end 
of the fission process. 
The variation of the neck generalized coordinate was
determined by using simple calculations of a minimal value of the
liquid drop energy. 

\begin{figure}
\includegraphics{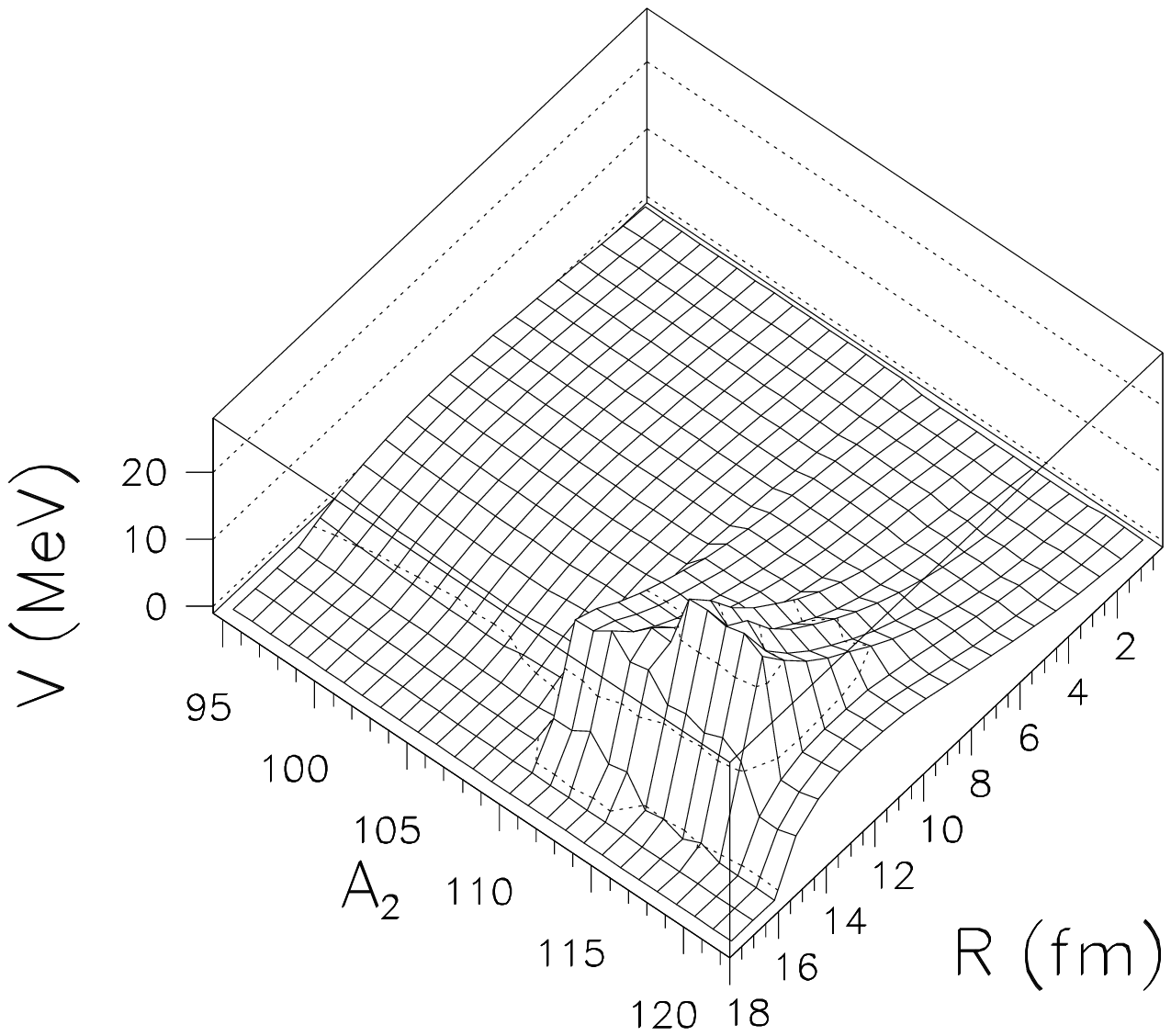}

\caption{Lichid drop potential energy $V_{LD}$ for the selected 
binary partitions
with respect to $A_{2}$ and the elongation $R$.
 }

\label{fig5}
\end{figure}

\begin{figure}
\includegraphics{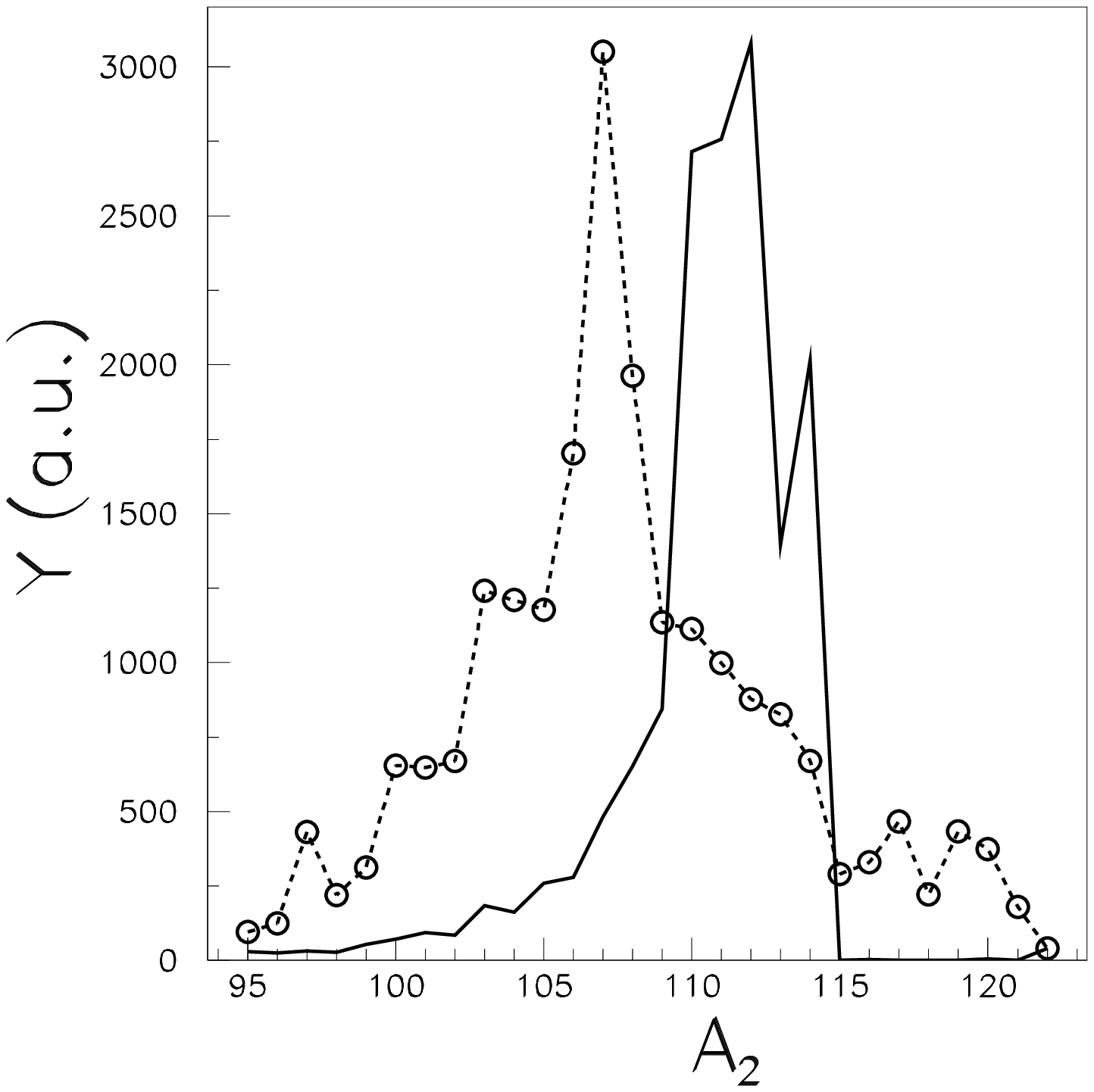}
\caption{Experimental yields in arbitrary units (dashed line),
compared with renormalized theoretical penetrabilities calculated within
the liquid drop model (solid line) with respect to $A_{2}$.
 }
\label{fig6}
\end{figure}

Dependencies of the liquid drop energy $V_{LD}$ as a
function of $C$ and $R$ are displayed in Fig. \ref{fig2} for two
nuclei, namely $^{236}$U and $^{252}$Cf. In these preliminary 
calculations, the deformations of the fragments are neglected. 
From the dependencies exhibited in these plots, 
it can be assessed that the influence of the
neck parameter is crucial, the variations of $V_{LD}$ exceeding several tens
of MeV for small shifts of $C$. For Cf, up to $R$=17 fm the liquid drop
deformation energy is always lower for negative values of $C$, that is for
swollen shapes. For values of $R>$17 fm, the necked shapes are favored and
the nucleus can be splitted into two separated fragments. 

From these simple considerations and by analyzing the macroscopic-
microscopic 
energy in the case of $^{252}$Cf for symmetric partitions, a particular
behavior  of $C$ as function of $R$ was extracted. For very 
low values of the 
elongation,  the values of $C$ are considered negative, 
so the nucleus is swollen in the median part. 
Up to $R<$ 8 fm, the curvature is varied linearly
up to the value $C$=0.4 fm$^-1$. From $R$=8 fm up to 
the exit from the second barrier the neck is kept positive.
The fact that the shapes are necked in the median surface
implies the formation two individual fragments during the fission
process itself, as expected in cold fission.
Thus, at 17 fm, a sudden change 
of the curvature is postulated, so that, the shapes become very necked 
in the median region and the two nuclei are separated. 
The variations of the generalized coordinates as function of the
elongation $R$ are plotted in Fig. \ref{fig3}. 
Dependencies between the generalized coordinates $q_{i}$ ($i=1,...5$)
supply the fission trajectory. The shapes obtained within
our nuclear shape parametrization are plotted in Fig. \ref{fig4},
starting from the ground state and reaching two separated 
deformed fragments. 

A complete description of the fission path requires
the minimization of the action integral. Such calculations are very
time consuming and they were performed up to now only for a single 
partition as in Refs. \cite{mir07,mir09}. 
In the actual calculations many channels have to be studied and such 
calculations are not feasible. To make our problem tractable, only one 
fission path is used for all investigated partitions.

\begin{figure}
\includegraphics{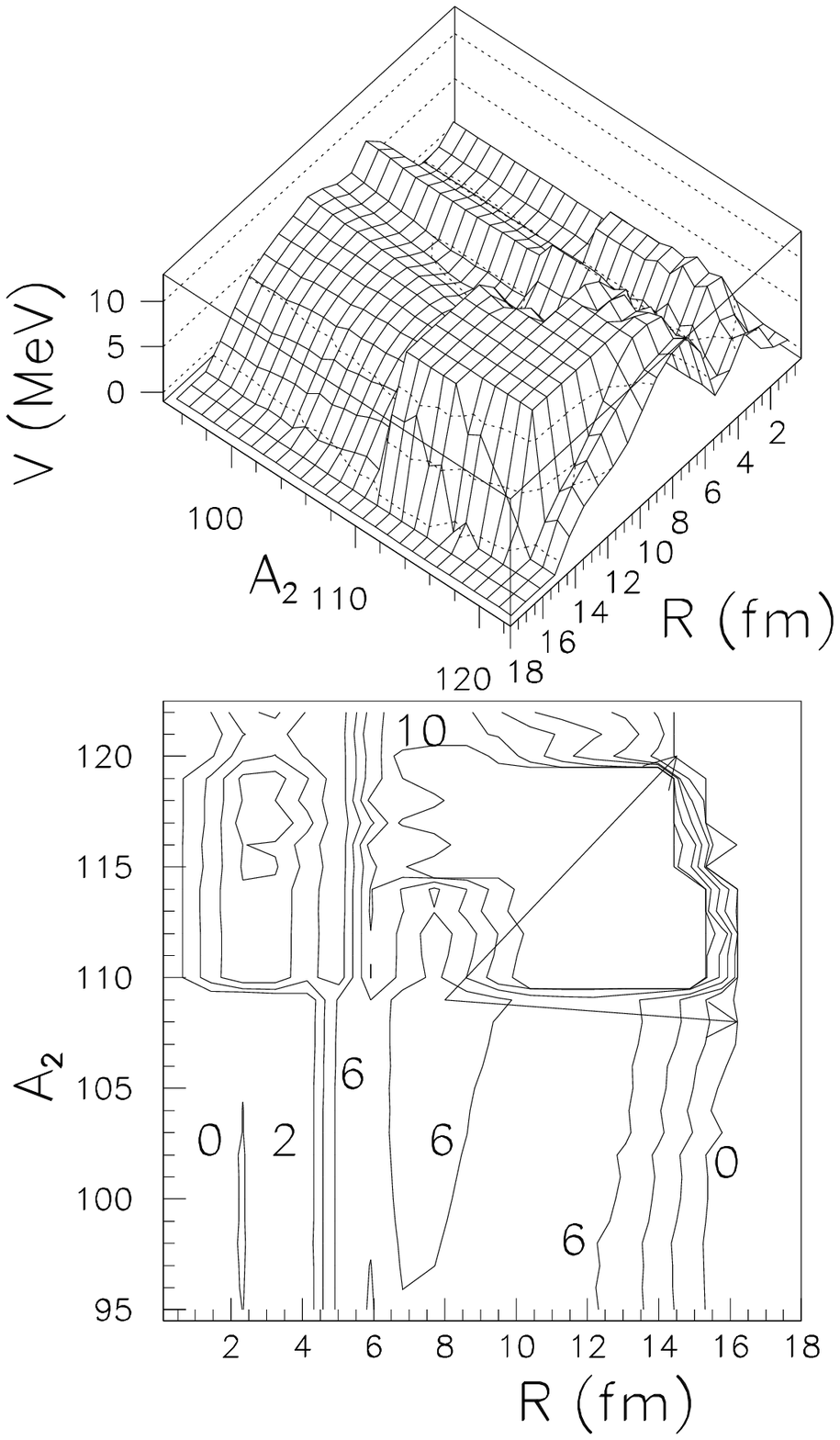}
\caption{Deformation energy $V$ computed within the microscopic-macroscopic
method for different binary partitions
with respect to $A_{2}$ and the elongation $R$.
In the lower plot, the equipotential contours are plotted in
step of 2 MeV. Some values of the deformation energy are marked.
 }
\label{fig7}
\end{figure}

\begin{figure}
\includegraphics{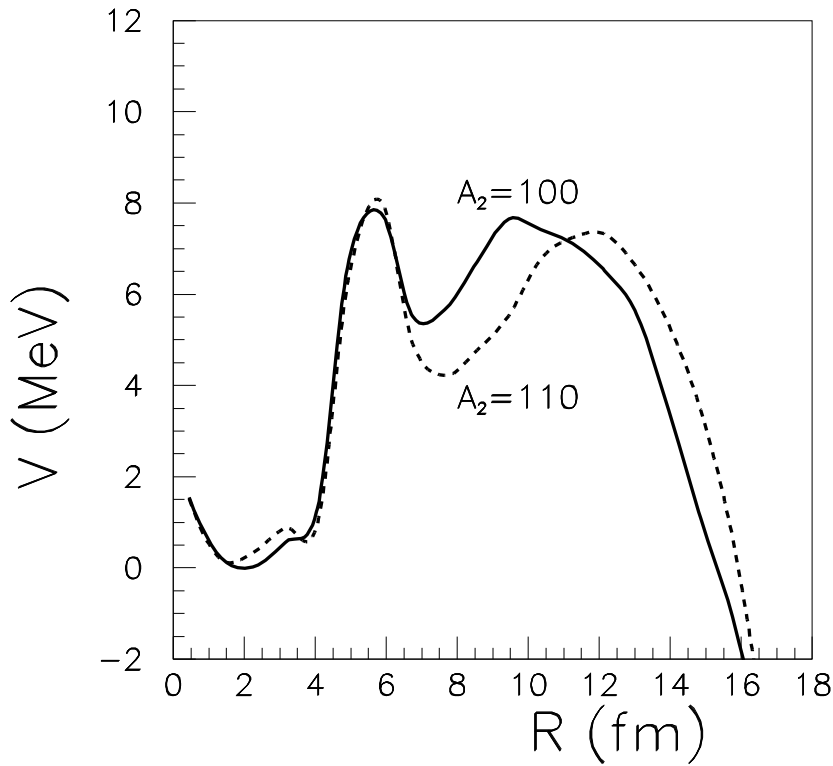}
\caption{Fission barriers obtained for two different partitions 
as function of the elongation. 
The mass of the light fragment is marked on the plot.
 }
\label{fig8}
\end{figure}

\begin{figure}
\includegraphics{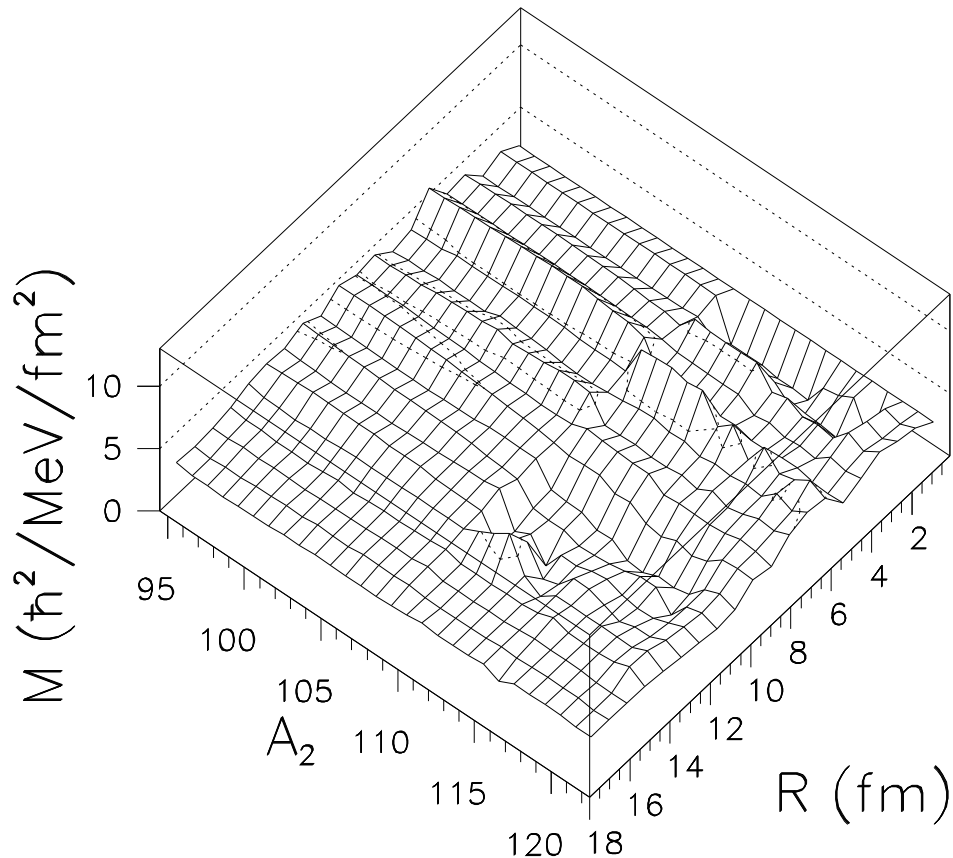}
\caption{Effective mass computed within the cranking approximation
for different binary partitions
with respect to $A_{2}$ and the elongation $R$.
 }
\label{fig9}
\end{figure}

\begin{figure}
\includegraphics{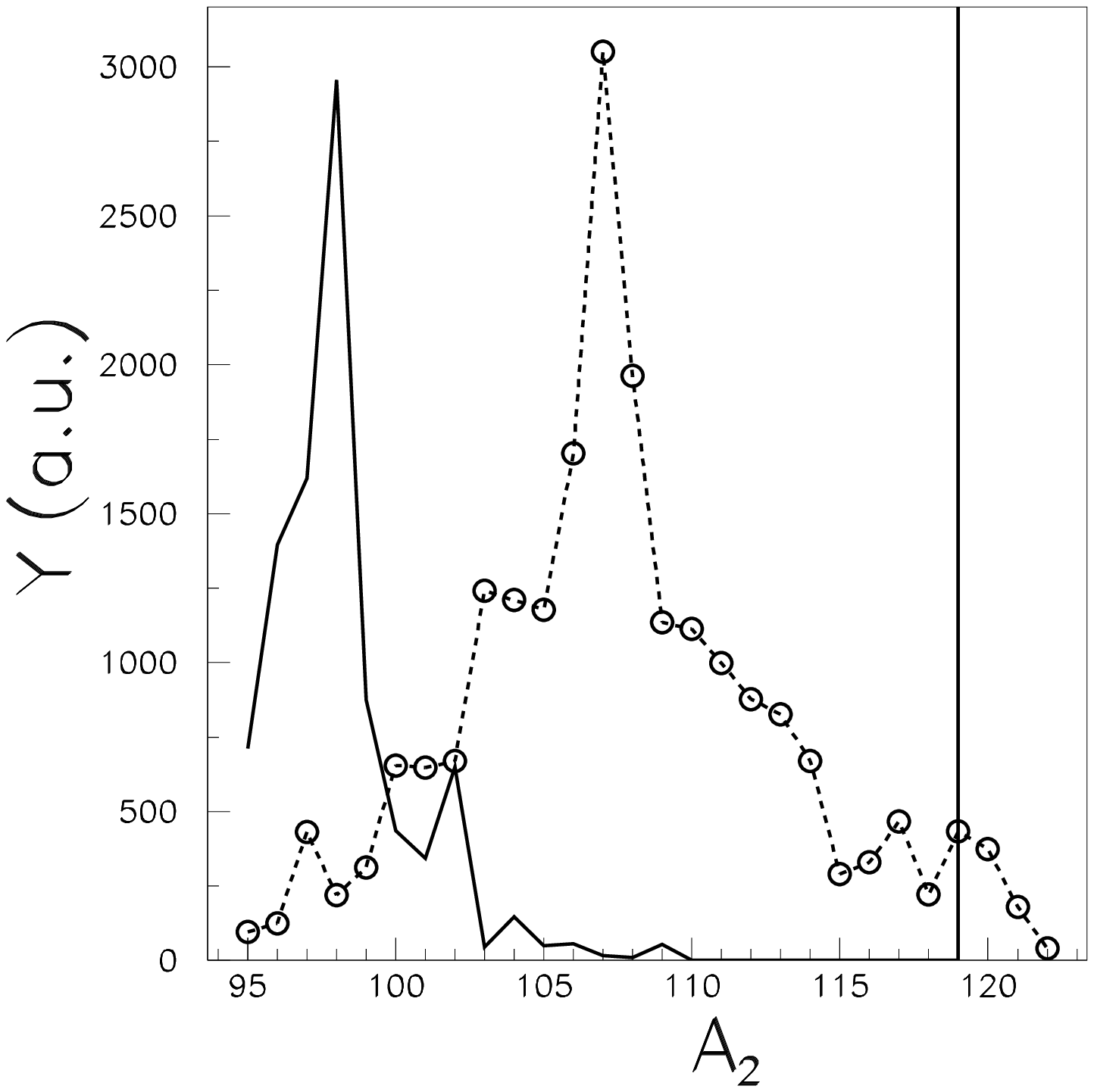}
\caption{Experimental yields in arbitrary units (dashed line),
compared with renormalized theoretical penetrabilities calculated within
the microscopic-macroscopic model (solid line) with respect to  $A_{2}$.
 }
\label{fig10}
\end{figure}

\begin{figure}
\includegraphics{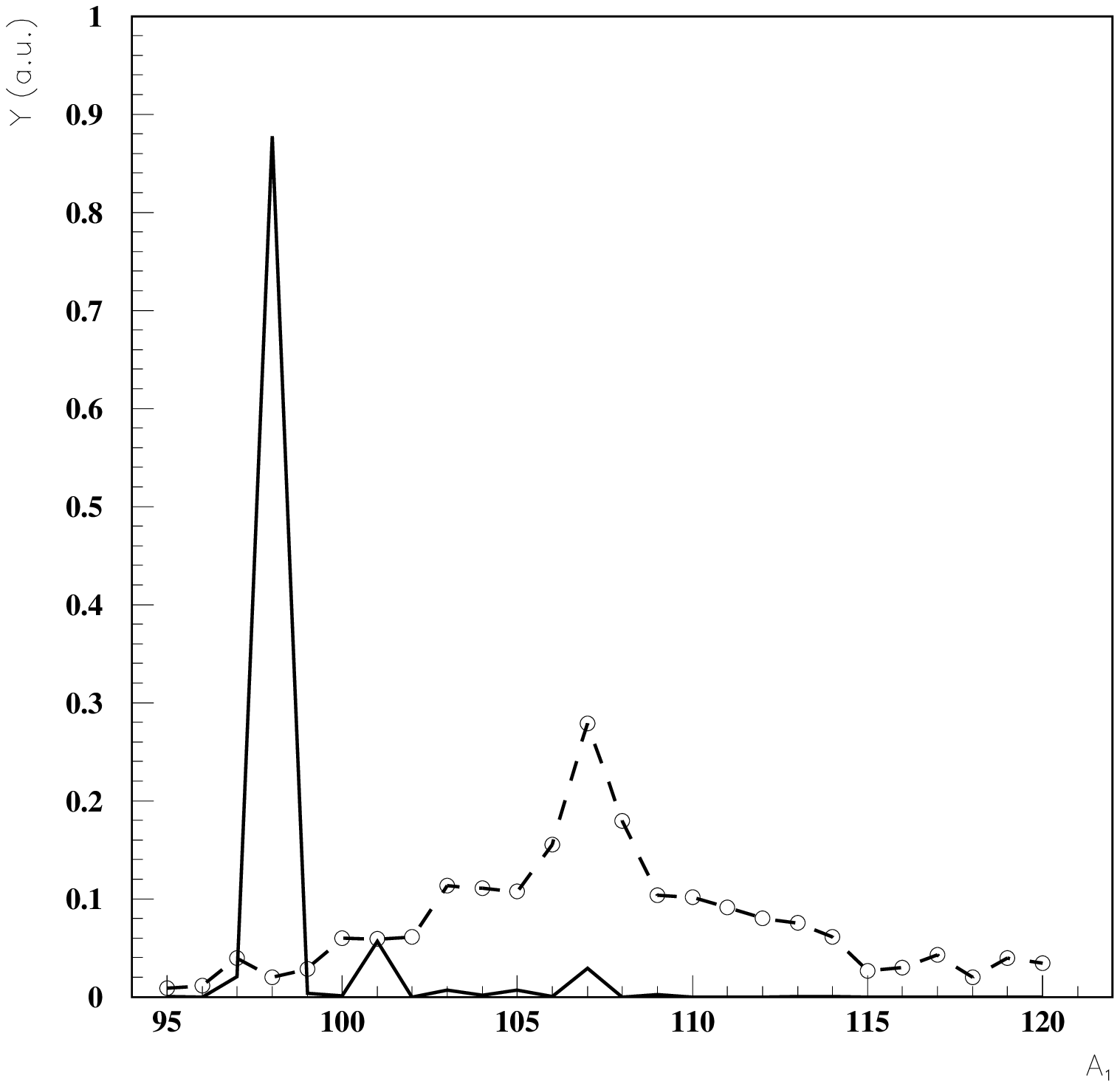}
\caption{Experimental yields in arbitrary units (dashed line),
compared with renormalized theoretical penetrabilities calculated within
the double folding model (solid line) with respect to  $A_{2}$.
 }
\label{fig10a}
\end{figure}

By using the previous parametrization, the liquid drop deformation
energy $V_{LD}$ is computed for all selected binary partitions. The results
are plotted in Fig. \ref{fig5} as functions of the light fragment 
$A_{2}$ and the elongation $R$. It is interesting to notice that the 
liquid drop deformation energy exhibits a strong increase for the
channels with $A_2>$110. This variation is caused by the change
of the ground state deformations of fission fragments, from prolate
shapes to oblate ones. A first evalution is realized for transmissions
of the liquid drop barrier.

\begin{figure}
\includegraphics{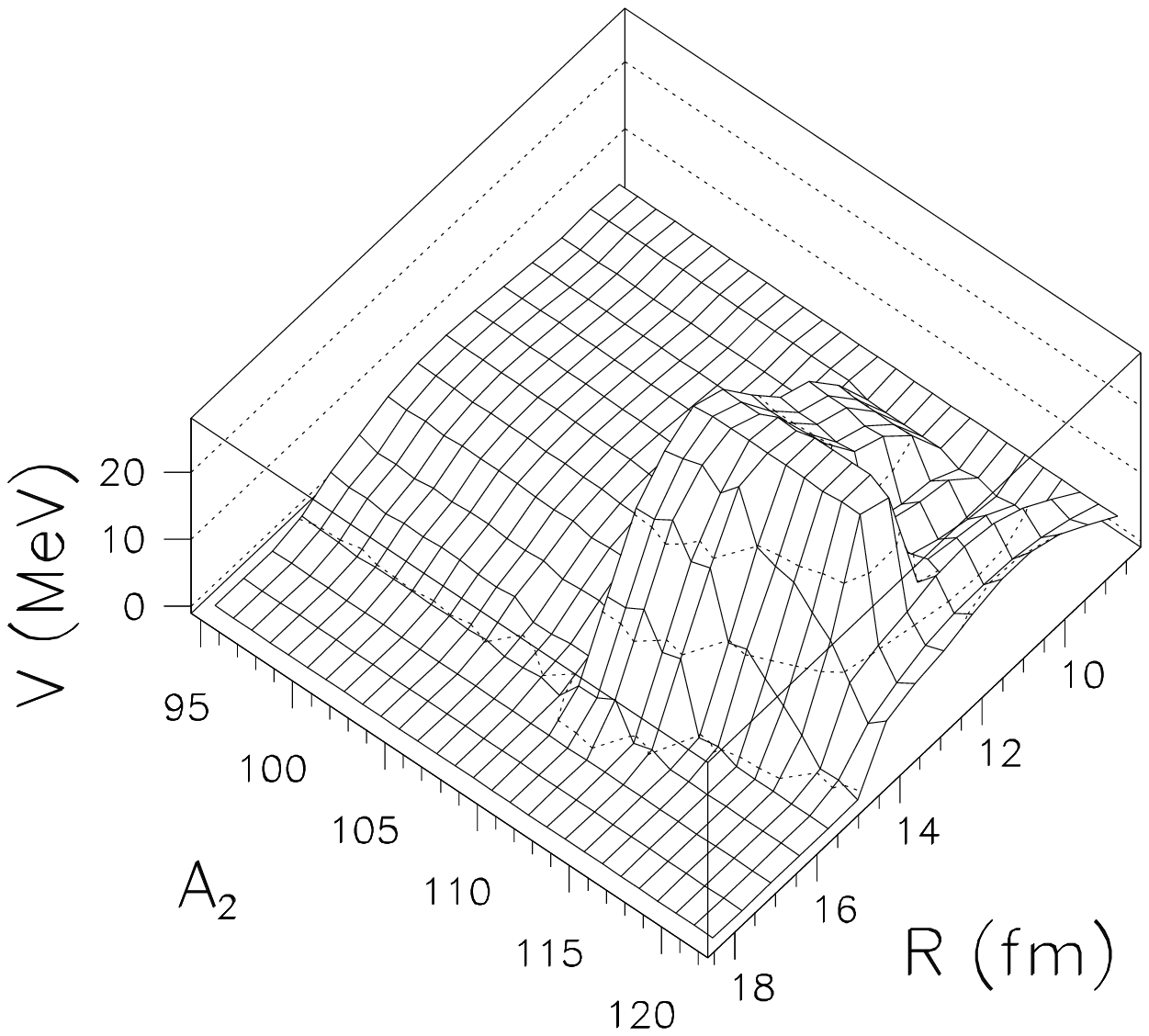}
\caption{Deformation energy $V$ calculated within the microscopic-macroscopic
method for different binary partitions
with respect to $A_{2}$ and the elongation $R$ beginning from
the minimum energy configuration of the second well and varying
linearly all generalized coordinates up to the scission point.
 }
\label{fig11}
\end{figure}

\begin{figure}
\includegraphics{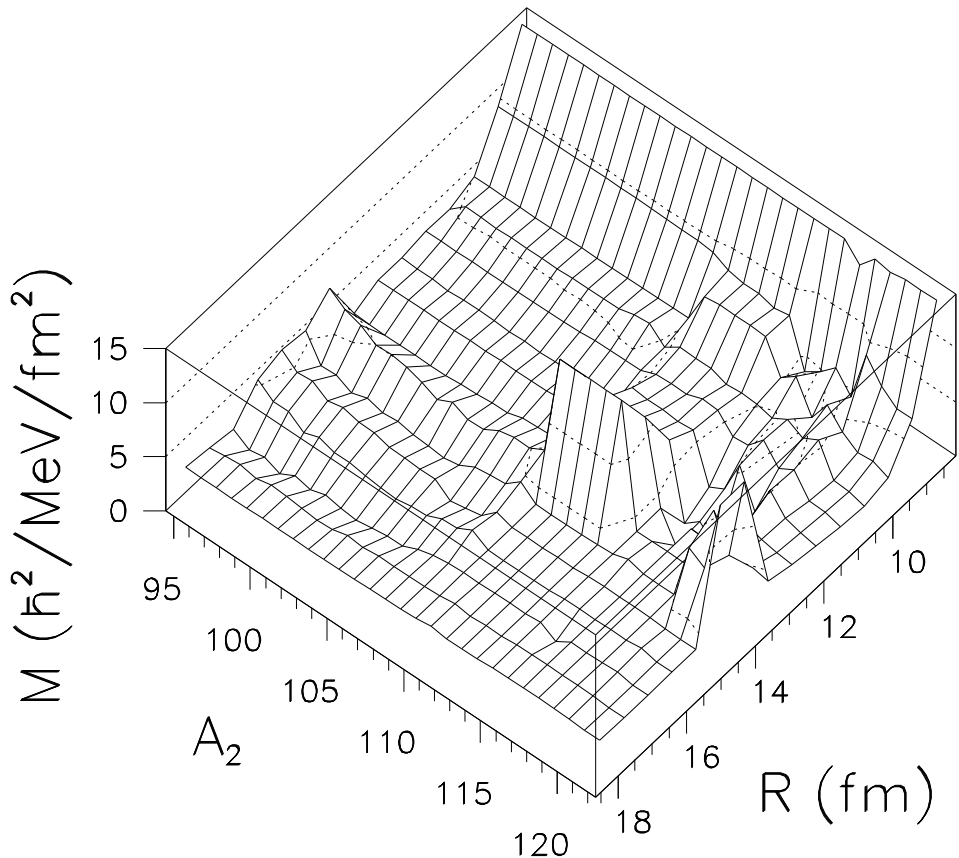}
\caption{Effective mass computed in the cranking approximation
for different binary partitions
with respect to $A_{2}$ and the elongation $R$ in the region of the
second barrier.
 }
\label{fig12}
\end{figure}

\begin{figure}
\includegraphics{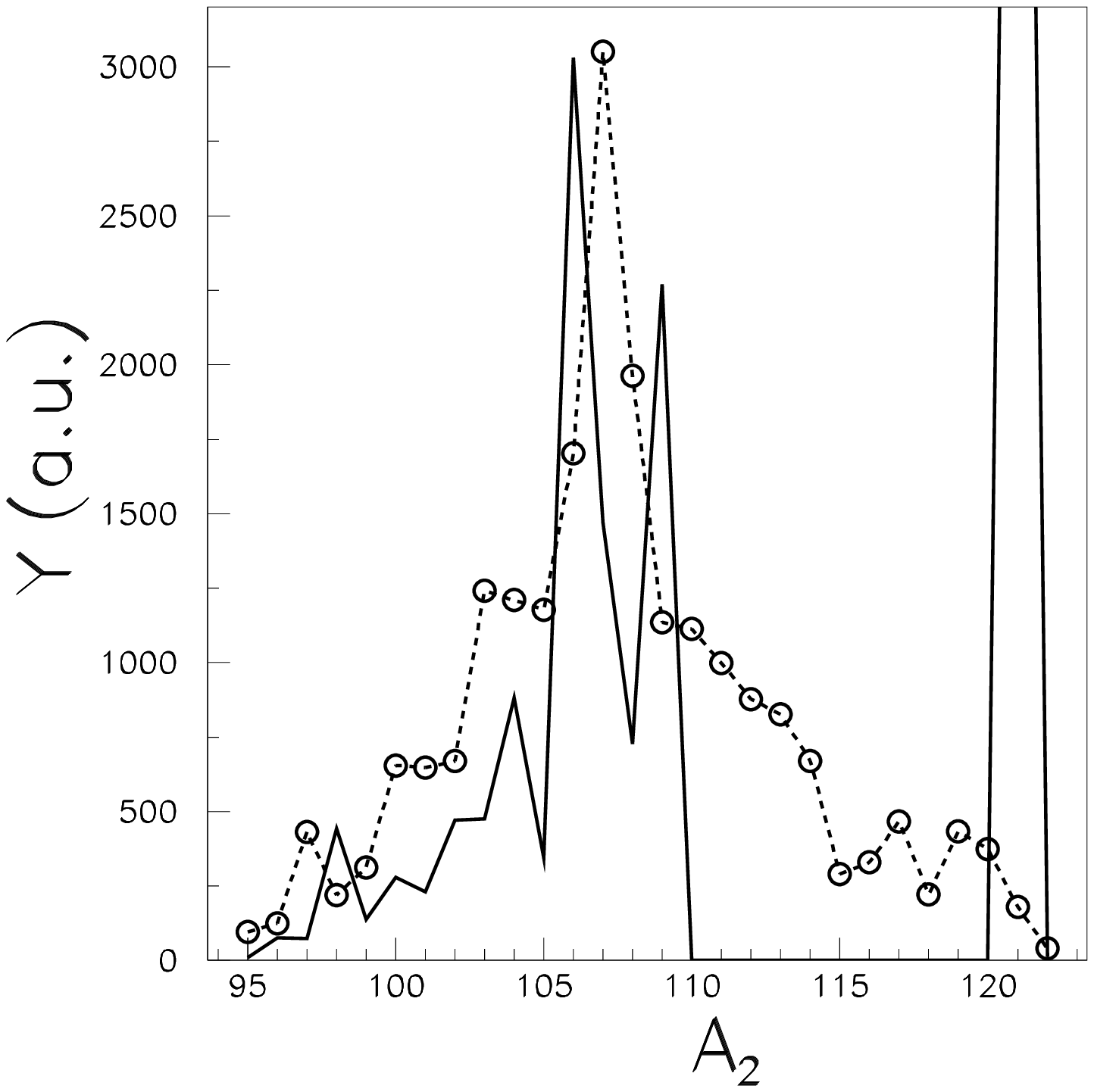}
\caption{Experimental yields in arbitrary units (dashed line),
compared  with renormalized theoretical penetrabilities calculated 
within the microscopic-macroscopic model (solid line)
with respect to  $A_{2}$ by taking into account the
minimum energy in the second well.
 }
\label{fig13}
\end{figure}

The penetrabilities are computed using the WKB approximation by considering
that the effective mass equals the reduced mass 
$\mu={A_1 A_2\over A}$ of each partition. 
The theoretical penetrabilities, renormalized such that 
the maximal value is comparable to the maximum experimental yields, are
plotted in Fig. \ref{fig6}. The maximal theoretical
yield tends to be located around mass 103. 
 The sudden drop
of the penetrability at $A_{2}$=110 is due to a change
in the deformations of the fragments. Up to $A_{2}$=110,
both fission fragments are prolate while after $A_{2}$=110
the deformations become oblate. The liquid drop model fails to reproduce
the experimental data. The maximum experimental yield is located at
$A_2$=
107 in the experimental data while the maximum theoretical
transmission is given for $A_2$=104. 

An improved description is expected by introducing shell and pairing 
effects in the framework of the microscopic-macroscopic model.
By using the same fission paths, the deformation energy $V$ 
is obtained in the framework of the microscopic-macroscopic model. 
The values of $V$
are plotted in Fig. \ref{fig7} as function of the elongation $R$ and
the light fragment mass $A_2$. Some general features of the fission
barrier can be extracted.
For $A_2 <$110 partitions, a double humped barrier is exhibited.
An interesting feature can be observed in
the region of the second well, 
where a global minimum for $A_2$=110 and
$R\approx$8 fm (the starting point of two arrows in the lower panel
of Fig. \ref{fig7}) is clearly evidenced. This minimum indicates
a possible isomeric state. In Fig. \ref{fig8}, two barriers 
corresponding to two different partitions with $A_{2}$=100 and 110 are 
plotted.

In both cases, the inner barriers have comparable values and shapes, but
the second minimum is lowered for higher $A_2$. On the other hand, 
the shapes of the external barriers are changed dramatically when
the mass-asymmetry varies. The external barrier is lower for
$A_2$=110 than for $A_2$=100. In the same time, the external
barrier is thinner for $A_2$=100. The competition of these two effects
are canceling when the transmission is computed.
In order to evaluate the influence of the deformation energy on the 
fragment yields, the action integral has to be be evaluated. 
In this context, the inertia is computed within the cranking 
approximation \cite{cranking} within the formula
\begin{eqnarray}
\mu=B_{RR}+B_{CC}\left({\partial C\over\partial R}\right)^{2}+
 B_{\eta\eta}\left({\partial \eta\over\partial R}\right)^{2}+
 B_{\epsilon_1\epsilon_1}\left({\partial \epsilon_1\over\partial R}\right)^{2}+
 B_{\epsilon_2\epsilon_2}\left({\partial \epsilon_2\over\partial R}\right)^{2}+\nonumber\\
2B_{C\eta}{\partial C\over\partial R}{\partial \eta \over \partial R}+
2B_{C\epsilon_1}{\partial C\over\partial R}{\partial \epsilon_1 \over \partial R}+
2B_{C\epsilon_2}{\partial C\over\partial R}{\partial \epsilon_2 \over \partial R}+\\
2B_{\eta\epsilon_1}{\partial \eta\over\partial R}{\partial \epsilon_1 \over \partial R}+
2B_{\eta\epsilon_2}{\partial \eta\over\partial R}{\partial \epsilon_2 \over \partial R}+
2B_{\epsilon_1\epsilon_2}{\partial \epsilon_1\over\partial R}{\partial 
\epsilon_2 \over \partial R}~,
\nonumber
\end{eqnarray}
where $B_{ij}$ are the elements of the effective mass tensor.
The resulting effective mass along the postulated trajectory
is plotted on Fig. \ref{fig9}. 
For comparison with experimental data, the theoretical
yields are considered proportional with the transmission.

The penetrabilities of the fission barrier,
obtained in the frame of the microscopic-macroscopic
model for zero excitation energy, are displayed in Fig. \ref{fig10}.
These penetrabilities reflect the interplay between the values
of the deformation energy and the inertia along a given fission path. 
Unfortunately, the theoretical
transmissions are not able to reproduce the main trends observed
in the experimental distribution of fission yields. By introducing
the shell effects, our results are even worse that those obtained within the
simple liquid drop framework. Two peaks
in the theoretical penetrabilities are observed at $A_2$=97 and 120,
while the maximum experimental yield is located at $A_2$=107.

Thus, a direct transmission of the barrier is not consistent with
experimental data. Therefore another effect should describe the yields 
distribution, this effect being neglected in our calculations.
It is worth mentioning that similar results are obtained 
in Fig. \ref{fig10a} by using the double folding procedure 
\cite{sandu98}, by using the same deformation parameters and neglecting 
the mass asymmetry.

It is possible that the cold fission follows another fission path
that takes into consideration the existence of a global minimum in the
second well, that is the cold fission trajectory must proceed through
the isomer configuration.
In these circumstances we consider that the
penetrability must be produced in two steps: first of all the isomeric
state in the second well is reached and afterwards, the second barrier
is penetrated. As previously mentioned, the isomeric state is located
at $A_2$=110 and $R\approx$ 8 fm. From this isomeric state, the nuclear shapes
and mass asymmetry are modified linearly such that at the exit
point of the barrier, the 
final shapes of the fragments are reached as displayed by the arrows 
plotted on the lower panel of Fig. \ref{fig7}. The isomer minimum
is considered axial symmetric.

The microscopic-macroscopic deformation energy starting only from
the isomeric state up to the formation of two separated fragments
is plotted in Fig. \ref{fig11}. The generalized shape coordinates
change linearly from the isomeric configuration values to the final ones 
at the exit point of the external barrier.
The inertia computed in the framework
of the cranking model is displayed in Fig. \ref{fig12}. With
these ingredients, the calculated penetrabilities at zero
excitation energies through the external barrier are compared with
experimental yields on Fig. \ref{fig13}. In this last approximation
it is implicitly assumed that the penetration of the inner barrier 
is the same for all partitions, so that differences in the 
barrier transmission between channels are managed only by the external barrier.

A very good agreement between the theoretical penetrability distribution
and the experimental yields is obtained for $A_2<110$. The maximum theoretical
value is obtained at $A_2$=106 while the maximum experimental yield is found
at $A_{2}$=107. A sudden drop of theoretical penetrabilities is 
theoretically obtained for channels characterized by $A_2>110$. 
As previously noticed, this behavior is connected with the ground state 
shapes of fragments that become oblate for these channels.
Perhaps these oblate shapes are not the best configurations during 
the penetration of the barrier, and the final ground state oblate
configurations are obtained only after the exit from the fission barrier.
It is important to notice that, in order to reproduce the
experimental data, a crucial hypothesis was made, namely
the cold fission is considered intimately connected to the deformations
of the ground state of nascent fragments. In general, 
the fission process is more favorable for
prolate nuclear shapes and it is possible that our hypothesis is no
longer valid. Therefore, it can be expected that the fission process proceed
through prolate deformations of the fragments and the oblate ground 
state shapes are obtained only after the exit from the outer barrier.
An enhancement of the fission probability is obtained for
$A_2$=120. I must be noticed that experimentally, a special situation
is encountered in cold fission of $^{252}$Cf with one fragment being
the $^{132}$Sn \cite{gon,cle}. Here the yield shows an enhancement
for very exciation energies. Such a structure was observed only for
fragmentations close to $^{132}$Sn. There is an indication that 
can be a cluster mode fission. This enhancement was
also reproduced by our calculations.

\section{Conclusions}
\label{sec:concl}

Concluding, we computed the potential energy surface for different binary
combinations, giving the superheavy compound nucleus $^{252}_{98}$Cf,
by using the Two Center Shell Model.
We obtained a satisfactory agreement of the experimental yields
by considering variable mass and charge asymmetry beyond the first
barrier of the potential surface.
The cold fission is strongly connected with Z=50 magic number, because 
an enhancement of the fission probability 
 was obtained for $^{120}_{47}$Ag+$^{132}_{51}$Sb 
partition.
The low value of the deformation energy in the isomeric microscopic 
configuration reflects the magicity of the Sn proton number. 
It was evidenced that a good agreement with experimental data can be 
obtained only if the fission path proceeds through this isomeric 
configuration.
The final transmissions are managed only by the external barrier, showing a 
strong dependence on the final fragment shapes. Due to the fact
that prolate shapes are more favorable for the fission process, the 
maximal values of the yields are
shifted towards channels characterized by lower values of $A_2$. 
Thus, the cold fission process of $^{252}$Cf can be called Tin-like 
radioactivity, similar with with the Led-like radioactivity, 
corresponding to various heavy cluster emission processes.

\section{Acknowledgements}

This work was supported by the contract IDEI-119 of the
Romanian Ministry of Education and Research.

\end{document}